\documentclass[a4paper,12pt]{article}
\usepackage{amssymb}
\usepackage{latexsym}
\usepackage[dvips]{graphicx}
\usepackage{cite}
\newcommand{\be}{\begin{equation}}
\newcommand{\ee}{\end{equation}}
\newcommand{\bea}{\begin{eqnarray}}
\newcommand{\nn}{\nonumber}
\newcommand{\eea}{\end{eqnarray}}

\begin{document}

\begin{titlepage}
\begin{flushright}
UB-ECM-PF-03/07
\end{flushright}
\begin{centering}
\vspace{.3in}
{\Large{\bf Energy Distribution in 2d Stringy Black Hole
Backgrounds}}
\\

\vspace{.5in} {\bf  Elias C.
Vagenas\footnote{evagenas@ecm.ub.es} }\\

\vspace{0.3in}

Departament d'Estructura i Constituents de la Mat\`{e}ria\\
and\\ CER for Astrophysics, Particle Physics and Cosmology\\
Universitat de Barcelona\\
Av. Diagonal 647\\ E-08028 Barcelona\\
Spain\\
\end{centering}

\vspace{0.7in}
\begin{abstract}
 We utilize M{\o}ller's and Einstein's energy-momentum complexes in order to
 explicitly evaluate the energy distributions associated with the two-dimensional
 ``Schwarzschild'' and ``Reissner-Nordstr\"{o}m'' black hole
 backgrounds. While M{\o}ller's prescription provides meaningful
 physical results, Einstein's prescription fails to do so
 in the aforementioned gravitational backgrounds. These
 results hold for all two-dimensional static black hole
 geometries. The results obtained within this context are exploited in order
 Seifert's hypothesis to be investigated.
\end{abstract}

\end{titlepage}
\newpage

\baselineskip=18pt
\section*{Introduction}
One of the most interesting problems which remains unsolved since
the birth of General Theory of Relativity, is the energy-momentum
localization. Many renowned physicists have been working on this
problematic issue with Einstein to be first in the row. After
Einstein's seminal work \cite{einstein} on energy-momentum
complexes a large number of expressions for the energy
distribution were proposed \cite{tolman,pp,ll,berg,gold,weinberg}.
These expressions were restricted to evaluate energy distribution
in quasi-Cartesian coordinates. M{\o}ller \cite{moller} proposed a
new expression for an energy-momentum complex which could be
utilized to any coordinate system. However, the idea of the
energy-momentum complex was severely criticized for a number of
reasons. Firstly, although a symmetric and locally conserved
object, its nature is nontensorial and thus its physical
interpretation seemed obscure  \cite{chandra}. Secondly, different
energy-momentum complexes could yield different energy
distributions for the same gravitational background
\cite{bergqvist1,bergqvist2}. Thirdly, energy-momentum complexes
were local objects while there was commonly believed that the
proper energy-momentum of the gravitational field was only total,
i.e. it cannot be localized \cite{chiang}. For a long time,
attempts to deal with this problematic issue were made only by
proposers of quasi-local approach \cite{brown,sean}.\par\noindent
In 1990 the concept of energy-momentum complexes and their use for
explicitly evaluating the energy distribution in given
gravitational backgrounds was revived by Virbhadra
\cite{vir1,vir2,vir3}. At the same time Bondi \cite{bondi}
sustained that a nonlocalizable form of energy is not admissible
in relativity so its location can in principle be found. Since
then, numerous works on evaluating the energy distribution of
several gravitational backgrounds have been completed employing
the abandoned for a long time approach of energy-momentum
complexes
\cite{par1,bak,par2,vir15,vir6,vir7,yang1,xulu1,xulu2,rad2,yang2,
xulu3,xulu4,rad1,xulu5,rad4,rad5,rad3,brin,xulu6}. In 1996
Aguirregabiria, Chamorro and Virbhadra \cite{vir4} showed that
five different\footnote{Later on Virbhadra \cite{vir5} came  to
know that Tolman's and Einstein's complexes which had been used in
\cite{vir4} were exactly the same (see footnote 1 in
\cite{vir5}).} energy-momentum complexes yield the same energy
distribution for any Kerr-Schild class metric. Additionally their
results were identical with the results of Penrose \cite{pen} and
Tod \cite{tod} using the notion of quasi-local mass. In 1999
Chang, Nester and Chen \cite{nester} proved that every
energy-momentum complex is associated with a Hamiltonian boundary
term. Thus, the energy-momentum complexes are quasi-local and
acceptable.
\par\noindent
In this paper we evaluate the energy distribution in
two-dimensional static black hole backgrounds utilizing the
Einstein's and M{\o}ller's prescriptions. Our interest in the
two-dimensional gravitational backgrounds stems to the fact that
lower dimensional theories of gravity provide simplified contexts
in which to study various physical phenomena \cite{2d}. The choice
of Einstein's approach among others was due to a recent work of
Virbhadra in which he points out that the description of Einstein
is the best one \cite{vir5}. The reasons for presenting here the
M{\o}ller's description are: (a) the argument that it is not
restricted to quasi-Cartesian coordinates and (b) a work of
Lessner \cite{les} who argues that the M{\o}ller's energy-momentum
complex is a powerful concept of energy and momentum in General
Theory of Relativity.
\par\noindent The remainder of the
paper is organized as follows. In Section 1 we present the concept
of energy-momentum complexes in the framework of General Theory of
Relativity. In Section 2 the expressions of the Einstein's
energy-momentum complexes in four and two dimensional
gravitational backgrounds are given. It is shown that for the case
of two-dimensional static gravitational backgrounds the
energy-momentum complex of Einstein cannot give meaningful
physical results. In Section 3 we present the M{\o}ller's
energy-momentum complexes in four and two dimensional
gravitational backgrounds. An explicit expression for the
energy-momentum complex of M{\o}ller in a two-dimensional static
gravitational background is derived. In Section 4 we use the
M{\o}ller's prescription in order to obtain the energy
distribution in a two-dimensional ``Schwarzschild'' black hole
background \cite{witten,mandal}. The results extracted in three
different coordinate systems (gauges) for the energy distribution
in the above-mentioned gravitational background are identical.  In
Section 5 the energy distribution in a two-dimensional
``Reissner-Nordstr\"{o}m'' black hole \cite{nappi,lee} in
M{\o}ller's prescription is explicitly calculated in
``Schwarschild'' gauge. The Seifert's hypothesis \cite{seifert} is
addressed in both Sections 4 and 5. Finally, Section 6 is devoted
to a brief summary of results and concluding remarks.
\section{Energy-Momentum Complexes}
The conservation laws of energy and momentum  for an isolated
(closed), i.e. no external force acting on the system, physical
system in the Special Theory of Relativity are expressed by a set
of differential equations. Defining $T^{\mu}_{\nu}$ as the
symmetric energy-momentum tensor of matter and all
non-gravitational fields the conservation laws are given by \be
T^{\mu}_{\nu,\, \mu} \equiv \frac{\partial T^{\mu}_{\nu}}{\partial
x^{\mu}}=0\ee where \be
\rho=T^{t}_{t}\hspace{1cm}j^{i}=T^{i}_{t}\hspace{1cm}p_{i}=-T^{t}_{i}\ee
are the energy density, the energy current density, the momentum
density, respectively, and Greek indices run over the spacetime
labels while Latin indices run over the spatial coordinate values.
\par\noindent
Making the transition from the Special to General Theory of
Relativity one adopts a simplicity principle which is called
principle of minimal gravitational coupling. As a result of this,
the conservation equation is now written as \be T^{\mu}_{\nu;\,
\mu} \equiv \frac{1}{\sqrt{-g}}\frac{\partial}{\partial
x^{\mu}}\left(\sqrt{-g}\,T^{\mu}_{\nu}\right)-\Gamma^{\kappa}_{\nu\lambda}T^{\lambda}_{\kappa}=0
\ee where $g$ is the determinant of the metric tensor
$g_{\mu\nu}(x)$. The conservation equation may also be written as
\be \frac{\partial}{\partial
x^{\mu}}\left(\sqrt{-g}\,T^{\mu}_{\nu}\right)=\xi_{\nu}\ee where
\be \xi_{\nu}=\Gamma^{\kappa}_{\nu\lambda}T^{\lambda}_{\kappa}\ee
is a nontensorial object. For $\nu=t$ this means that the matter
energy is not a conserved quantity for the physical
system\footnote{It is possible to restore the conservation law by
introducing a local inertial system for which at a specific
spacetime point $\xi_{\nu}=0$ but this equality by no means holds
in general.}. From a physical point of view this lack of energy
conservation can be understood as the possibility of transforming
matter energy into gravitational energy and vice versa. However,
this remains a problem and it is widely believed that in order to
be solved one has to take into account the gravitational energy
\cite{tolman,ll,weinberg,moller} .\par\noindent By a well-known
procedure, the nontensorial object $\xi_{\nu}$ can be written  as
\be \xi_{\nu}=-\frac{\partial}{\partial
x^{\mu}}\left(\sqrt{-g}\,\vartheta^{\mu}_{\nu}\right)\ee where
$\vartheta^{\mu}_{\nu}$ are certain functions of the metric tensor
and its first order derivatives. Therefore, the energy-momentum
tensor of matter $T^{\mu}_{\nu}$ is replaced by the expression \be
\theta^{\mu}_{\nu}=\sqrt{-g}\left(T^{\mu}_{\nu}+\vartheta^{\mu}_{\nu}\right)\ee
which is called energy-momentum complex since it is a combination
of the tensor $T^{\mu}_{\nu}$ and a pseudotensor
$\vartheta^{\mu}_{\nu}$ which describes the energy and  momentum
of the gravitational field. The energy-momentum complex satisfies
a conservation law in the ordinary sense, i.e. \be
\theta^{\mu}_{\nu,\, \mu}=0\ee and it can be written as \be
\theta^{\mu}_{\nu}=\chi^{\mu\lambda}_{\nu,\,\lambda}\ee where
$\chi^{\mu\lambda}_{\nu}$ are called superpotentials and are
functions of the metric tensor and its first order derivatives.
\par\noindent It is obvious that the energy-momentum complex is not
uniquely determined by the condition that is usual divergence is
zero since it can always been added to the energy-momentum complex
a quantity with an identically vanishing divergence.
\section{Einstein's Prescription}
The energy-momentum complex of Einstein in a  four-dimensional
background is given as \cite{einstein} \be
\theta^{\mu}_{\nu}=\frac{1}{16\pi}h^{\mu\lambda}_{\nu\,\, ,\,
\lambda}\label{etheta}\ee where the Einstein's superpotential $
h^{\mu\lambda}_{\nu}$ is of the form \be
h^{\mu\lambda}_{\nu}=\frac{1}{\sqrt{-g}}
g_{\nu\sigma}\left[-g\,\left(g^{\mu\sigma}g^{\lambda\kappa}\,
-\,g^{\lambda\sigma}g^{\mu\kappa}\right)\right]_{,\kappa}\label{esuper}\ee
with the antisymmetric property\be
h^{\mu\lambda}_{\nu}=-h^{\lambda\mu}_{\nu}\hspace{1ex}.\ee Thus,
the energy and momentum in Einstein's prescription for a
four-dimensional background are given by \be P_{\mu}=\int\int\int
\theta^{0}_{\mu}dx^{1}dx^{2}dx^{3}\ee and specifically the energy
of the physical system in a four-dimensional background is \be
E=\int\int\int
\theta^{0}_{0}dx^{1}dx^{2}dx^{3}\hspace{1ex}.\label{eenergy}\ee It
should be noted that the calculations have to be restricted to the
use of quasi-Cartesian coordinates.\par\noindent In the case of
two-dimensional gravitational backgrounds we have to modify
expressions (\ref{etheta}) and (\ref{eenergy}). Therefore, the
energy-momentum complex for a physical system in Einstein's
prescription for a two-dimensional gravitational background is
given as \be\theta^{\mu}_{\nu}=\frac{1}{4}h^{\mu\lambda}_{\nu\,\,
,\, \lambda}\label{etheta1}\ee and the energy is of the form \be
E=\int \theta^{0}_{0}dx^{1}\label{eenegry1}\hspace{1ex}\ee while
the expression (\ref{esuper}) for the  superpotential remains the
same.\par\noindent If we are interested to evaluate the energy of
a physical system in a two-dimensional gravitational background
which is static, then the superpotentials $h^{0\lambda}_{0}$ are
easily calculated \bea h^{00}_{0}&=&\frac{1}{\sqrt{-g}}
g_{00}\left[-g\,\left(g^{00}g^{01}\,
-\,g^{00}g^{01}\right)\right]_{,1}\\
h^{01}_{0}&=&\frac{1}{\sqrt{-g}}
g_{00}\left[-g\,\left(g^{00}g^{11}\,
-\,g^{10}g^{01}\right)\right]_{,1}\eea and after imposing
staticity we get
\bea h^{00}_{0}&=&0\\
h^{01}_{0}&=&\frac{1}{\sqrt{-g}}
g_{00}\left[-g\,g^{00}g^{11}\right]_{,1}=0\hspace{1ex}. \eea
Therefore, for the case of two-dimensional static gravitational
backgrounds the energy in Einstein's prescription will always be
\bea E&=&\frac{1}{4}\int h^{0\lambda}_{0\, ,\lambda}dx^{1}\nn\\&=&
\frac{1}{4}\int \left(h^{00}_{0\, ,0}+h^{001}_{0\,
,1}\right)dx^{1}\nn\\
&=&0\hspace{1ex}.\eea It is obvious that the Einstein's
prescription cannot be used in order to extract the energy
distribution associated with two-dimensional static gravitational
backgrounds.
\section{M{\o}ller's Prescription}
The energy-momentum complex of M{\o}ller in a  four-dimensional
background is given as \cite{moller}\be
\mathcal{J}^{\mu}_{\nu}=\frac{1}{8\pi}\xi^{\mu\lambda}_{\nu\,\,
,\, \lambda}\label{mtheta}\ee where the M{\o}ller's superpotential
$ \xi^{\mu\lambda}_{\nu}$ is of the form \be
\xi^{\mu\lambda}_{\nu}=\sqrt{-g} \left(\frac{\partial
g_{\nu\sigma} }{\partial x^{\kappa} }-\frac{\partial
g_{\nu\kappa}}{\partial x^{\sigma}
}\right)g^{\mu\kappa}g^{\lambda\sigma}\label{msuper}\ee with the
antisymmetric property\be
\xi^{\mu\lambda}_{\nu}=-\xi^{\lambda\mu}_{\nu}\hspace{1ex}.\ee
Thus, the energy and momentum in M{\o}ller's prescription for a
four-dimensional background are given by \be P_{\mu}=\int\int\int
\mathcal{J}^{0}_{\mu}dx^{1}dx^{2}dx^{3}\ee and specifically the
energy of the physical system in a four-dimensional background is
\be E=\int\int\int
\mathcal{J}^{0}_{0}dx^{1}dx^{2}dx^{3}\hspace{1ex}.\label{menergy}\ee
\par\noindent Since we will be interested in
two-dimensional gravitational backgrounds we have to modify
expressions (\ref{mtheta}) and (\ref{menergy}). Therefore, the
energy-momentum complex for a physical system in M{\o}ller's
prescription for a two-dimensional gravitational background is
given as
\be\mathcal{J}^{\mu}_{\nu}=\frac{1}{2}\xi^{\mu\lambda}_{\nu\,\,
,\, \lambda}\label{mtheta1}\ee and the energy is of the form \be
E=\int \mathcal{J}^{0}_{0}dx^{1}\label{menegry1}\hspace{1ex}\ee
while the expression (\ref{msuper}) for M{\o}ller's superpotential
remains the same.
\par\noindent
It should be noted that the calculations are not anymore
restricted to quasi-Cartesian coordinates but they can be utilized
in any coordinate system.
\par\noindent If we are interested to evaluate the energy of
a physical system in a two-dimensional gravitational background
which is static, then the M{\o}ller's superpotentials
$\xi^{0\lambda}_{0}$ are easily calculated \bea
\xi^{00}_{0}&=&\sqrt{-g} \left(\frac{\partial g_{0\sigma}
}{\partial x^{\kappa} }-\frac{\partial g_{0\kappa}}{\partial
x^{\sigma}
}\right)g^{0\kappa}g^{0\sigma}\\
\xi^{01}_{0}&=&\sqrt{-g} \left(\frac{\partial g_{0\sigma}
}{\partial x^{\kappa} }-\frac{\partial g_{0\kappa}}{\partial
x^{\sigma} }\right)g^{0\kappa}g^{1\sigma}\eea and after imposing
staticity we get
\bea \xi^{00}_{0}&=&0\\
\xi^{01}_{0}&=&-\sqrt{-g} \frac{\partial g_{00} }{\partial x^{1}
}\,g^{00}g^{11}\label{msuper011}\hspace{1ex}. \eea Therefore, for
the case of two-dimensional static gravitational backgrounds the
energy in M{\o}ller's prescription will always be \bea
E&=&\frac{1}{2}\int \xi^{0\lambda}_{0\, ,\lambda}dx^{1}\nn\\&=&
\frac{1}{2}\int \left(\xi^{00}_{0\, ,0}+\xi^{01}_{0\,
,1}\right)dx^{1}\nn\\
&=&\frac{1}{2}\int \left[-\left(\sqrt{-g} \frac{\partial g_{00}
}{\partial x^{1}
}\,g^{00}g^{11}\right)_{,1}\right]dx^{1}\label{menergy11}\hspace{1ex}.\eea
It is obvious that the M{\o}ller's prescription can provide us
with meaningful results for the energy distribution associated
with two-dimensional static gravitational backgrounds.
\section{``Schwarzschild'' Black Hole}
We start with the action in two spacetime dimensions
\cite{callan}\be S=\frac{1}{2\pi}\int
d^{2}x\sqrt{-g}\,e^{-2\phi}\left[R+4\left(\nabla\phi\right)^{2}+\lambda^{2}\right]
\label{action}\ee where $g$ is the determinant of the metric
$g_{\mu\nu}(x)$ in two spacetime dimensions, $\phi$ is the dilaton
field and $\lambda^{2}$ is the cosmological constant. This action
arises as the effective action describing the radial modes of
extremal dilatonic black holes in four or higher
dimensions\cite{gibbons1,gibbons2,ghs1,ghs2,gid}. The black hole
solution of (\ref{action}) was given by E. Witten \cite{witten} as
the low-energy approximation to an exact solution of string
theory. The line element of the above-mentioned stringy
two-dimensional ``Schwarzschild'' black hole solution can be
written in different gauges (coordinate systems) as follows
\cite{elias}:
\newline {\bf i) ``Schwarzschild" gauge}
\newline
The two-dimensional dilatonic black hole in the ``Schwarzschild"
gauge is characterized by the line element: \be
ds^2=-g(r)dt^2+g^{-1}(r)dr^2 \label{schdil} \ee where the function
$g(r)$ is given by: \be g(r)=1-\frac{M}{\lambda}e^{-2\lambda
r}\label{metric} \ee and $0<t<+\infty$, $r_H<r<+\infty$, with $
r_H=\frac{1}{2\lambda}ln(\frac{M}{\lambda})$ the position of the
event horizon of the black hole.
\par\noindent
Following the terminology of the previous section, the covariant
 components of the metric function are \bea g_{00}&=&-
 \left(1-\frac{M}{\lambda}e^{-2\lambda r}\right)\label{sgd00}\\
 g_{11}&=&\frac{1}{\left(1-\frac{M}{\lambda}e^{-2\lambda
 r}\right)}\label{sgd11}\hspace{1ex}, \eea
 the corresponding contravariant
components are \bea
g^{00}=-\frac{1}{\left(1-\frac{M}{\lambda}e^{-2\lambda r}\right)}\label{sgu00}\\
g^{11}=\left(1-\frac{M}{\lambda}e^{-2\lambda
r}\right)\label{sgu11}\eea and the determinant of the metric
function $ g_{\mu\nu}(x)$ is \be g=-1\label{sdet}\hspace{1ex}.\ee
In order to evaluate the energy distribution in M{\o}ller's
prescription  associated with the exterior of the two-dimensional
``Schwarzschild'' black hole we evaluate the nonzero
superpotential (\ref{msuper011}) \be \xi^{01}_{0}=-2M e^{-2\lambda
r }\ee and substituting it in
 equation (\ref{menergy11}) we get \bea E_{{\tiny ext}}&=&\frac{1}{2}\int^{+\infty}_{R}
 \left[-\left(2M e^{-2\lambda
r }\right)_{,r}\right]dr\nn\\&=&-M e^{-2\lambda r}
 \Bigg| ^{+\infty}_{R}\nn\\&=& M e^{-2\lambda R}
 \label{sext}\hspace{1ex}.\eea
Since it has been proved that the asymptotic value of the total
gravitational mass of a two-dimensional ``Schwarzschild'' black
hole is the mass parameter $M$ which appear in the metric function
(\ref{metric}) it is clear that the energy associated with a
two-dimensional ``Schwarzschild'' black hole in a sphere of radius
$R$ will be \be E (R)=M\left(1-e^{-2\lambda R}\right)
\label{sdistri}\hspace{1ex}.\ee There are some comments in order.
Firstly, the energy of two-dimensional ``Schwarzschild'' black
hole is distributed to its interior as well as its exterior.
Secondly, since the cosmological constant $\lambda$ is positive,
the energy distribution is positive everywhere, even in the
forbidden region, i.e. $0<r<r_{H}$. Therefore, a neutral test
particle in the aforementioned gravitational background
experiences at a finite radial distance a positive effective
gravitational mass, given by equation (\ref{sdistri}). Thirdly,
Seifert conjectured \cite{seifert} that any singularity that
occurs is hidden if a finite nonzero amount of matter tends to
collapse into one point, or is naked if either one has
singularities along lines (or surfaces) or the central
singularities are carefully arranged that they contain zero mass.
It is clear from equation (\ref{sdistri}) that at the origin, i.e.
$R\!=\!0$, the mass of the two-dimensional ``Schwarzschild'' black
hole is equal to zero providing a support to Seifert's hypothesis.
\newline
{\bf ii) Unitary gauge}
\newline
The line element is: \be ds^2=-tanh^2(\lambda y)dt^2+dy^2
\label{unidil} \ee where the ``unitary" variable $y$ is given by
the following expression: \be
y=\frac{1}{\lambda}ln\left[e^{\lambda (r-r_H)} + \sqrt{e^{2\lambda
(r-r_H)}-1}\right] \ee and $0<y<+\infty$.
\par\noindent
Following the terminology of Section 3, the covariant
 components of the metric function are \bea g_{00}&=&-tanh^2(\lambda y)
 \label{ugd00}\\
 g_{11}&=&1\label{ugd11}\hspace{1ex}, \eea
 the corresponding contravariant
components are \bea
g^{00}&=&-\frac{1}{tanh^2(\lambda y)}\label{ugu00}\\
g^{11}&=&1\label{ugu11}\eea and the determinant of the metric
function $ g_{\mu\nu}(x)$ is now given by\be g=-tanh^2(\lambda
y)\label{udet}\hspace{1ex}.\ee In order to evaluate the energy
distribution in M{\o}ller's prescription associated with the
exterior of the two-dimensional ``Schwarzschild'' black hole but
in the unitary gauge now, we evaluate the nonzero superpotential
(\ref{msuper011}) \be
\xi^{01}_{0}=-\frac{2\lambda}{cosh^{2}(\lambda y)}\ee and
substituting it in
 equation (\ref{menergy11}) we get \bea E_{{\tiny
 ext}}&=&\frac{1}{2}\int^{+\infty}_{R}
 \left[-\left(\frac{2\lambda}{cosh^{2}(\lambda
 y)}\right)_{,y}\right]dy\nn\\
 &=&\lambda \left[tanh^{2}(\lambda y)- 1\right]\Bigg| ^{+\infty}_{R}
 \label{uext}\hspace{1ex}.\eea  It is easily seen that since equations
 (\ref{sgd00}) and (\ref{ugd00}) are equal, the energy
 distribution in the unitary gauge (\ref{uext}) associated with the exterior
 of a two-dimensional ``Schwarzschild'' black hole is equal to
 the corresponding energy distribution derived in the Schwarzschild gauge (\ref{sext})
 and therefore the energy associated with a
two-dimensional ``Schwarzschild'' black hole in a sphere of radius
$R$ in the unitary gauge will be \be E (R)=M\left(1-e^{-2\lambda
R}\right) \label{udistri}\ee which is the same, as expected, with
the corresponding expression (\ref{sdistri}) derived in the
``Schwarzschild'' gauge. \par\noindent Obviously, the comments
made in the ``Schwarzschild'' gauge also hold here.
\newline
{\bf iii) Conformal gauge}
\newline
The line element in this gauge is: \be ds^2=(1+e^{-2\lambda
x})^{-1}(-dt^2+dx^2) \label{condil} \ee where the variable $x$ is
given by: \be
x=\frac{1}{2\lambda}ln\left[e^{2\lambda(r-r_H)}-1\right]\label{ccoord}
\ee and $-\infty<x<+\infty$. \par\noindent Following as before the
terminology of the Section 3, the covariant
 components of the metric function are \bea g_{00}&=&-
 \frac{1}{\left(1+e^{-2\lambda
x}\right)}\label{cgd00}\\
 g_{11}&=&\frac{1}{\left(1+e^{-2\lambda
x}\right)}\label{cgd11}\hspace{1ex}, \eea
 the corresponding contravariant
components are \bea g^{00}&=&-\left(1+e^{-2\lambda
x}\right)\label{cgu00}\\
g^{11}&=&\left(1+e^{-2\lambda x}\right)\label{cgu11}\eea and the
determinant of the metric function $ g_{\mu\nu}(x)$ is \be
g=-\frac{1}{\left(1+e^{-2\lambda
x}\right)^{2}}\label{cdet}\hspace{1ex}.\ee In order to evaluate
the energy distribution in M{\o}ller's prescription associated
with the exterior of the two-dimensional ``Schwarzschild'' black
hole we evaluate the nonzero superpotential (\ref{msuper011}) \be
\xi^{01}_{0}=-\frac{2\lambda e^{-2\lambda x}}{\left(1+e^{-2\lambda
x}\right)}\ee and substituting it in
 equation (\ref{menergy11}) we get \bea E_{{\tiny ext}}&=&
 \frac{1}{2}\int^{+\infty}_{R}
 \left[-\left(\frac{2\lambda e^{-2\lambda x}}{\left(1+e^{-2\lambda
x}\right)}\right)_{,x}\right]dx\nn\\\nn\\&=&-\frac{\lambda
e^{-2\lambda x}}{\left(1+e^{-2\lambda
x}\right)}\label{cext}\hspace{1ex}.\eea After some, but not
lengthy, calculations it is easily seen, using equations
(\ref{ccoord}), (\ref{cgd00}) and the equality of the latter with
(\ref{sgd00}), that the energy
 distribution in the conformal gauge (\ref{cext}) associated with the exterior
 of a two-dimensional ``Schwarzschild'' black hole is equal to
 the corresponding energy distribution derived in the Schwarzschild gauge (\ref{sext})
 and therefore the energy associated with a
two-dimensional ``Schwarzschild'' black hole in a sphere of radius
$R$ in the conformal gauge will be \be E (R)=M\left(1-e^{-2\lambda
R}\right) \label{cdistri}\ee which is the same, as expected, with
the corresponding expressions (\ref{sdistri}) and (\ref{udistri})
derived in the ``Schwarzschild'' and unitary gauge, respectively.
\par\noindent Obviously, the comments made in the
``Schwarzschild'' gauge also hold here.
\section{``Reissner-Nordstr\"{o}m'' Black Hole}
Our starting point in this section will be  a two-dimensional
effective action realized in heterotic string theory
\cite{nappi}\be S=\frac{1}{2\pi}\int d^{2}x
\sqrt{-g}e^{-2\phi}\left[R+4(\nabla\phi)^{2}+4\lambda^{2}-\frac{1}{2}F^{2}\right]\label{caction}\ee
where $g$ is the determinant of the two-dimensional metric
$g_{\mu\nu}(x)$, $\phi$ is the dilaton, $\lambda^{2}$ is the
cosmological constant and $F_{\mu\nu}$ is the Maxwell stress
tensor.\par\noindent The line element of the charged
two-dimensional black hole solution derived from (\ref{caction})
is given by (in coordinates corresponding to the ``Schwarzschild"
gauge of the previous section) \cite{lee}: \be ds^2 = -g(r)dt^2 +
g^{-1}(r)dr^2 \label{linelement2} \ee where \be g(r) = 1 -
\frac{M}{\lambda} e^{-2\lambda r} + \frac{Q^2}{4 \lambda ^2}
e^{-4\lambda r} \label{rnmetric} \ee with $0<t<+\infty$,
$r_+<r<+\infty$, $r_+$ being the future event horizon of the black
hole. The parameters $M$ and $Q$ are the mass and electric charge
, respectively, of the ``Reissner-Nordstr\"{o}m'' black hole
(\ref{linelement2}-\ref{rnmetric}) which is the corresponding
charged black hole for the two-dimensional ``Schwarzschild'' black
hole described in the previous section.
\par\noindent
Following the terminology of the Section 3, the covariant
 components of the metric function are \bea g_{00}&=&-\left(1 -
\frac{M}{\lambda} e^{-2\lambda r} + \frac{Q^2}{4 \lambda ^2}
e^{-4\lambda r}\right)
 \label{rngd00}\\
 g_{11}&=&\left(1 -
\frac{M}{\lambda} e^{-2\lambda r} + \frac{Q^2}{4 \lambda ^2}
e^{-4\lambda r}\right)^{-1}\label{rngd11}\hspace{1ex}, \eea
 the corresponding contravariant
components are \bea g^{00}=-\left(1 - \frac{M}{\lambda}
e^{-2\lambda r} + \frac{Q^2}{4 \lambda ^2}
e^{-4\lambda r}\right)^{-1}\label{rngu00}\\
g^{11}=\left(1 - \frac{M}{\lambda} e^{-2\lambda r} + \frac{Q^2}{4
\lambda ^2} e^{-4\lambda r}\right)\label{rngu11}\eea and the
determinant of the metric function $ g_{\mu\nu}(x)$ is \be
g=-1\label{rndet}\hspace{1ex}.\ee In order to evaluate the energy
distribution in M{\o}ller's prescription  associated with the
exterior of the two-dimensional ``Reissner-Nordstr\"{o}m'' black
hole we calculate the nonzero superpotential (\ref{msuper011}) \be
\xi^{01}_{0}=-2Me^{-2\lambda r }+\frac{Q^{2}}{\lambda}e^{-4\lambda
r }\ee and substituting it in
 equation (\ref{menergy11}) we get \bea E_{{\tiny ext}}
 &=&\frac{1}{2}\int^{+\infty}_{R}\left[-\left(2Me^{-2\lambda r }-\frac{Q^{2}}{\lambda}e^{-4\lambda
r }\right)_{,r}\right]\nn\\&=&\left[-M e^{-2\lambda r}
 +\frac{Q^2}{2\lambda}e^{-4\lambda r}\right]\Bigg| ^{+\infty}_{R}\nn\\&=& M e^{-2\lambda R}
 -\frac{Q^2}{2\lambda}e^{-4\lambda R}\label{nrext}\hspace{1ex}.\eea
It has been shown that the asymptotic value of the total
gravitational mass of a two-dimensional ``Reissner-Nordstr\"{o}m''
black hole is mass parameter $M$ which appear in the metric
function (\ref{rnmetric}). Therefore, it is obvious that the
energy associated with a two-dimensional
``Reissner-Nordstr\"{o}m'' black hole in a sphere of radius $R$
will be \be E (R)=M - M e^{-2\lambda
R}+\frac{Q^2}{2\lambda}e^{-4\lambda R}
\label{rndistri}\hspace{1ex}.\ee It is understood that the energy
of the two-dimensional ``Reisnner-Nordstr\"{o}m'' black hole is
distributed to its interior as well as its exterior. Switching off
the electric charge, i.e. $Q=0$, we get that equation
(\ref{rndistri}) goes over to (\ref{sdistri}).  Additionally,
treating $\lambda$ as a positive constant, the energy distribution
is positive everywhere, even in the forbidden region, i.e.
$0<r<r_{H}$ and therefore a neutral test particle in the
aforementioned gravitational background experiences at a finite
radial distance a positive effective gravitational mass, given by
equation (\ref{rndistri}). Finally, it is easily seen from
equation (\ref{rndistri}) that at the origin, i.e. $R\!=\!0$, the
energy of the two-dimensional ``Reissner-Nordstr\"{o}m'' black
hole is nonzero and this is a counterexample\footnote{The case
$Q=0$ has been disregarded here since this is the
``Schwarzschild'' case.} to Seifert's hypothesis \cite{seifert}.
\section{Conclusions}
In this work, we explicitly calculate the energy distributions
associated with the two-dimensional ``Schwarzschild'' and
``Reissner-Nordstr\"{o}m'' black holes. We, firstly, obtain the
exact expressions of M{\o}ller's and Einstein's energy-momentum
complexes in two-dimensional gravitational backgrounds. Although
M{\o}ller's and Einstein's approaches associated with four
dimensional gravitational backgrounds give meaningful results, in
the case of two-dimensional static gravitational backgrounds
Einstein's complex is proved to be always zero\footnote{The fact
that Einstein's prescription fails to yield meaningful physical
results in two-dimensional static gravitational backgrounds should
not lead one to doubt about the validity of this approach in
general.}. Therefore, since the two-dimensional ``Schwarzschild''
and ``Reissner-Nordstr\"{o}m'' black holes are static backgrounds,
we are led to work only with M{\o}ller's energy-momentum complex.
We have found the explicit expression for the energy distribution
associated with a two-dimensional ``Schwarzschild'' black hole.
The result was extracted in three different coordinate systems
(gauges) and found to be the same in all three gauges, verifying
M{\o}ller's assertion that his expression of energy-momentum
complex can be applied in any coordinate system giving meaningful
outcomes. It is shown that the energy of the aforesaid stringy
background is distributed to its interior as well as its exterior.
Since this energy is positive everywhere, even behind the event
horizon, i.e. $r_H$, a neutral test particle in this gravitational
field ``feels'' a positive effective gravitational mass which is
the energy that we have derived in M{\o}ller's approach.
Concerning the two-dimensional ``Reissner-Nordstr\"{o}m'' black
hole analogous results are obtained. We calculate its energy
distribution working only in the ``Schwarzschild'' gauge.
Switching off the electric charge, i.e. $Q=0$, we get the same
result as in the case of the two-dimensional ``Schwarzschild''
black hole. Finally, it is noteworthy to observe that the energy
distributions associated with the  two-dimensional
``Schwarzschild'' and ``Reissner-Nordstr\"{o}m'' black hole
backgrounds provide  an example and a counterexample,
respectively, to the Seifert's hypothesis. We therefore agree with
Virbhadra about the necessity of an adequate prescription for
localization or quasi-localization of mass before discussing
useful hypotheses within the framework of General Theory of
Relativity.
\section*{Acknowledgements}
The author is indebted to Professor K.S. Virbhadra for useful
suggestions and comments on the manuscript. This work has been
supported by the European Research and Training Network
``EUROGRID-Discrete Random Geometries: from Solid State Physics to
Quantum Gravity" (HPRN-CT-1999-00161).


\begin{thebibliography}{99}
\bibitem{einstein} A. Einstein, Sitzungsber.Preuss.Akad.Wiss.Berlin
(Math.Phys.) 778 (1915), Addendum-ibid. 799 (1915).

\bibitem{tolman} R.C. Tolman, {\it Relativity, Thermodynamics and
Cosmology}, (Oxford University Press, London) 227 (1934).

\bibitem{pp} A. Papapetrou, Proc. R. Ir. Acad. A {\bf52}, 11 (1948).

\bibitem{ll} L.D. Landau and E.M. Lifshitz, {\it The Classical Theory of
Fields}, (Addison-Wesley Press, Reading, MA) 317 (1951).

\bibitem{berg} P.G. Bergmann and R. Thompson, Phys. Rev. {\bf89},
400 (1953).

\bibitem{gold} J.N. Goldberg, Phys. Rev. {\bf111}, 315 (1958).

\bibitem{weinberg} S. Weinberg, {\it Gravitation and Cosmology: Principles
and Applications of General Theory of Relativity}, (Wiley, New
York) 165 (1972).

\bibitem{moller} C. M{\o}ller, Ann. Phys. (N.Y.) {\bf4}, 347
(1958).

\bibitem{chandra} S. Chandrasekhar and V. Ferrari, Proc. R. Soc.
London A {\bf435}, 645 (1991).

\bibitem{bergqvist1} G. Bergqvist, Class. Quant. Grav. {\bf9}, 1753
(1992).

\bibitem{bergqvist2} G. Bergqvist, Class. Quant. Grav. {\bf9},
1917 (1992).

\bibitem{chiang} C.M. Chen and J.M. Nester, Class. Quant. Grav.
{\bf16}, 1279 (1999).

\bibitem{brown} J.D. Brown and J.W. York, Phys. Rev. D {\bf47},
1407 (1993).

\bibitem{sean} S.A. Hayward, Phys. Rev. D {\bf49}, 831 (1994).

\bibitem{vir1} K.S. Virbhadra, Phys. Rev. D {\bf41}, 1086 (1990).

\bibitem{vir2} K.S. Virbhadra, Phys. Rev. D {\bf42}, 1066 (1990).

\bibitem{vir3} K.S. Virbhadra, Phys. Rev. D {\bf42}, 2919 (1990).

\bibitem{bondi} H. Bondi, Proc. R. Soc. London A {\bf427}, 249
(1990).
\bibitem{par1} K.S. Virbhadra and J.C. Parikh, Phys. Lett. B
{\bf317}, 312 (1993).

\bibitem{bak} D. Bak, D. Cangemi and R. Jackiw, Phys. Rev. D
{\bf49}, 5173 (1994).

\bibitem{par2} K.S. Virbhadra and J.C. Parikh, Phys. Lett. B
{\bf331}, 302 (1994); Erratum-ibid B {\bf340}, 265 (1994).

\bibitem{vir15} K.S. Virbhadra, Pramana-J. Phys. {\bf45}, 215 (1995).

\bibitem{vir6} A. Chamorro and K.S. Virbhadra, {\it Energy of a spherically
symmetric charged dilaton black hole}, Published in Spanish
Relativity Meeting, gr-qc/9602005.

\bibitem{vir7} A. Chamorro and K.S. Virbhadra, Pramana-J. Phys. {\bf45}, 181 (1995).

\bibitem{yang1} I.C. Yang, W.F. Lin and R.R. Hsu, Chin. J. Phys.
{\bf37}, 118 (1999).

\bibitem{xulu1} S.S. Xulu, Int. J. Theor. Phys. {\bf37}, 1773 (1998).

\bibitem{xulu2} S.S. Xulu, Int. J. Mod. Phys. D {\bf7}, 773
(1998).

\bibitem{rad2} I. Radinschi, Acta Phys. Slov. {\bf49}, 789 (1999).


\bibitem{yang2} I.C. Yang, Chin. J. Phys. {\bf38}, 1040 (2000).


\bibitem{xulu3} S.S. Xulu, Int. J. Mod. Phys. A {\bf15}, 2979 (2000).

\bibitem{xulu4} S.S. Xulu, Int. J. Mod. Phys. A {\bf15}, 4849 (2000).

\bibitem{rad1} I. Radinschi, Mod. Phys. Lett. A {\bf15}, 803 (2000).

\bibitem{xulu5} S.S. Xulu, Mod. Phys. Lett. A {\bf15}, 1511 (2000).

\bibitem{rad4} I. Radinschi, Mod. Phys. Lett. A {\bf15}, 2171 (2000).

\bibitem{rad5} I. Radinschi, Mod. Phys. Lett. A {\bf16}, 673 (2001).

\bibitem{rad3} I.C. Yang and I. Radinschi, Mod. Phys. Lett. A {\bf17}, 1159
(2002).

\bibitem{brin} T. Bringley, Mod. Phys. Lett. A {\bf17}, 157 (2002).

\bibitem{xulu6} S.S. Xulu, Astrophys. Space Sci. {\bf283}, 23
(2003).
\bibitem{vir4} J.M. Aguirregabiria, A.Chamorro and K.S. Virbhadra,
Gen. Rel. Grav. {\bf28}, 1393 (1996).

\bibitem{pen} R. Penrose, Proc. R. Soc. London A {\bf381}, 53
(1982).

\bibitem{tod} K.P. Tod, Proc. R. Soc. London A {\bf388}, 457
(1983).

\bibitem{nester} C.C. Chang, J.M. Nester and C.M. Chen, Phys. Rev.
Lett. {\bf83}, 1897 (1999).

\bibitem{2d} J.A. Harvey and A. Strominger, {\it String Theory
and Quantum Gravity '92: Proc. Trieste Spring School \& Workshop},
(ICTP, 1992) ed. J. Harvey et al (World Scientific, Singapore).

\bibitem{vir5} K.S. Virbhadra, Phys. Rev. D {\bf60}, 104041 (1999).

\bibitem{les} G. Lessner, Gen. Rel. Grav. {\bf28}, 527 (1996).

\bibitem{witten} E. Witten, Phys. Rev. D {\bf44}, 314 (1991).

\bibitem{mandal} G. Mandal, A.M. Sengupta and S.R. Wadia,
Mod. Phys. Lett. A {\bf6},  1685 (1991).

\bibitem{nappi} M.D. McGuigan, C.R. Nappi and S.A. Yost, Nucl.
Phys. B {\bf375}, 421 (1992).

\bibitem{lee} H.W. Lee, Y.S. Myung, J.Y. Kim and D.K. Park,
Class. Quant. Grav. {\bf14}, L53 (1997).

\bibitem{seifert} H.J. Seifert, Gen. Rel. Grav. {\bf10},
1065 (1979).

\bibitem{callan} C.G. Callan, S.B. Giddings, J.A. Harvey and A.
Strominger, Phys. Rev. D {\bf45}, R1005 (1992).

\bibitem{gibbons1} G.W. Gibbons, Nucl. Phys. B {\bf207}, 337
(1982).

\bibitem{gibbons2} G.W. Gibbons and K. Maeda, Nucl. Phys. B {\bf298},
741 (1988).

\bibitem{ghs1} D. Garfinkle, G. Horowitz and A. Strominger, Phys.
Rev. D {\bf43}, 3140 (1991).

\bibitem{ghs2} G. Horowitz and A. Strominger, Nucl. Phys. B {\bf360},
197 (1991).

\bibitem{gid} S.B. Giddings and A. Strominger, Phys. Rev. Lett.
{\bf67}, 2930 (1991).

\bibitem{elias} T. Christodoulakis, G.A. Diamandis, B.C. Georgalas
and E.C. Vagenas, Phys. Lett. B {\bf501}, 269 (2001).







\bibitem{adm} R. Arnowitt, S. Deser and C.W. Misner, in {\it Gravitation: An Introduction to Current
Research}, ed. by L. Witten (Wiley, New York, 1962).


 \end{thebibliography}
\end{document}